\documentclass[preprint]{aastex63}
\newcommand{\beq}{\begin{equation}}
\newcommand{\eeq}{\end{equation}}
\newcommand{\beqn}{\begin{eqnarray}}
\newcommand{\eeqn}{\end{eqnarray}}

\accepted{May 22, 2022}

\shorttitle{Particle Acceleration in Relativistic Shearing Flows}
\shortauthors{Rieger and Duffy}

\begin{document}

\title{Particle Acceleration in Relativistic Shearing Flows: Energy Spectrum}

\correspondingauthor{Frank M. Rieger}
\email{f.rieger@uni-heidelberg.de}

\author[0000-0003-1334-2993]{Frank M. Rieger}
\affiliation{Institute for Theoretical Physics, University of Heidelberg, 
Philosophenweg 16, D-69120 Heidelberg, Germany}
\affiliation{Max-Planck-Institut f\"ur Kernphysik, Saupfercheckweg 1, D-69117 Heidelberg, Germany}

\author{Peter Duffy}
\affiliation{School of Physics, University College Dublin, Belfield, Dublin 4, Ireland}

\nocollaboration{2}



\begin{abstract}
We consider the acceleration of charged particles in relativistic shearing flows, with Lorentz 
factor up to $\Gamma_0 \sim 20$. We present numerical solutions to the particle transport equation 
and compare these with results from analytical calculations. We show that in the highly relativistic 
limit the particle energy spectrum that results from acceleration approaches a power law, 
$N(E)\propto E^{-\tilde{q}}$, with a universal value $\tilde{q}=(1+\alpha)$ for the slope of this 
power law, where $\alpha$ parameterizes the power-law momentum dependence of the particle mean 
free path. At mildly relativistic flow speeds, the energy spectrum becomes softer and sensitive 
to the underlying flow profile. We explore different flow examples, including Gaussian and 
power-law-type velocity profiles, showing that the latter yield comparatively harder spectra, 
producing $\tilde{q}\simeq 2$ for $\Gamma_0 \simeq 3$ and Kolmogorov turbulence. We provide a 
comparison with a simplified leaky-box approach and derive an approximate relation for estimating 
the spectral index as a function of the maximum shear flow speed.
These results are of relevance for jetted, high-energy astrophysical sources such as active galactic 
nuclei, since shear acceleration is a promising mechanism for the acceleration of charged particles 
to relativistic energies and is likely to contribute to the high-energy radiation observed.
\end{abstract}

\section{Introduction}
The relativistic outflows from gamma-ray bursts and active galactic nuclei (AGNs) are generically expected
to exhibit velocity shearing across the flow. In AGNs, for example, lateral shearing could be intrinsically 
(jet-wind topology) or extrinsically (interaction with the environment, matter entrainment) induced.
The growing realization that such shear flows can be conducive to efficient Fermi-type particle acceleration
\citep[e.g.,][]{Lemoine2019,Rieger2019} has in recent times motivated a variety of studies exploring its 
role for the production of high-energy particles and emission \citep[e.g.,][]{Webb2018, Webb2019, Webb2020,
Rieger2019, Rieger2019b, Rieger2021,Tavecchio2021,Merten2021, Wang2021}. 
Early models of particle transport in shearing flows implied that in the absence of radiative losses the
accelerated particle (momentum) distribution in steady state follows a power-law spectrum, $f(p) \propto 
p^{-s}$, with spectral index $s=(3+\alpha)$ for a mean scattering time scaling as $\tau \propto p^{\alpha}$
\citep{Berezhko1981,Berezhko1982,Rieger2006}. 
It was subsequently shown, however, that this index only applies to the highly relativistic ($\beta\rightarrow1$) 
limit \citep{Webb2018,Webb2019,Rieger2019b}. In particular, subrelativistic flow speeds lead to much softer 
spectra. The reason for this is that in diffusive shear acceleration the momentum gain, for a particle scattered 
across the flow, competes with diffusive escape from the system \citep{Jokipii1990,Rieger2006} thereby 
affecting the particle spectrum. 

The leaky box model for the spatially averaged, steady-state  momentum phase-space distribution $f(p)$ 
captures these essential features \citep{Rieger2019b}. With a momentum diffusion coefficient $D_p$, 
escape timescale $\tau_{\rm esc}$ and injection $Q(p)$ this equation is 
\beq\label{diffusion}
\frac{1}{p^2} \frac{\partial}{\partial p }\left(p^2 D_p \frac{\partial f}{\partial p} \right) 
- \frac{f}{\tau_{\rm esc}}
+ Q(p) = 0 \,,
\eeq in the regime where radiative losses are negligible.
With a momentum-dependent mean free path, $\lambda\propto p^\alpha$, the diffusion coefficient 
$D_p\propto p^{2+\alpha}$ leads to an acceleration timescale that decreases with momentum 
$t_{\rm acc}(p)=p^2/([4+\alpha] D_p)\propto p^{-\alpha}$ as higher-momentum particles are scattered 
across a greater part of the shear flow. The same scattering gives rise to a spatial diffusion 
coefficient $\kappa(p)\propto\lambda\propto p^\alpha$ and an escape timescale $\tau_{\rm esc}=
(\Delta r)^2/2\kappa(p)\propto p^{-\alpha}$ where $\Delta r$ is the width of the shear region. That 
the acceleration and escape timescales have the same momentum dependence is a key requirement 
of any Fermi-type acceleration process to result in a power-law distribution. With $f\propto p^{-s}$ 
inserted into Equation (\ref{diffusion}), the index is 
\beq\label{s_LB}
 s = \frac{(3+\alpha)}{2} + \sqrt{\frac{(3+\alpha)^2}{4} + (4+\alpha)\, \frac{t_{\rm acc}}{\tau_{\rm esc}}}\,.
 \eeq
recovering the limit of $s=3+\alpha$ when $t_{\rm acc}\ll \tau_{\rm esc}$. 

The full transport equation for the phase-space density $f(r,p)$ includes both momentum and spatial 
diffusion terms with a spatially dependent flow velocity $c\beta(r)$. In principle, this will change 
the prediction for the spectral index in the leaky-box limit of equation (\ref{s_LB}). 
The aim of this paper is to calculate spectral indices for the full transport equation and to examine 
how the details of the flow profile affect the value for $s$. 
To achieve this, we compare in the following (Sec.~2) numerical solutions of the relativistic particle 
transport equation for three major key flow profiles, and we discuss them in the context of this 
simplified leaky-box approach in Sec.~3. 
The results provide a foundation for connecting momentum power-law spectral indices to specific 
flow velocity profiles.

\section{Particle Transport in Relativistic Flows}
The phase-space evolution of energetic charged particles in relativistic flows obeys a diffusive 
particle transport equation \citep[e.g.,][]{Webb1989}. In the case of a relativistic jet with 
cylindrical shear flow $\beta(r)=u_z(r)/c$, the corresponding, steady-state particle transport 
equation for the phase-space distribution function $f(r,p)$ with source term $Q$ takes the form 
\citep{Webb2018}
\beq\label{pte}
\frac{1}{p^2}
\frac{\partial}{\partial p} \left(p^4 \Gamma_s \tau \frac{\partial f}{\partial p}\right) 
+\frac{1}{r} \frac{\partial}{\partial r} \left(\kappa r \frac{\partial f}{\partial r}\right)
+ Q=0\,,
\eeq where $\tau = \lambda/c$ is the mean scattering time, $\kappa =c^2 \tau / 3$ is the spatial diffusion
 coefficient, and 
\beq\label{GammaS}
\Gamma_s \equiv \Gamma_s(r) = \frac{c^2}{15}\, \gamma_b(r)^4 \left(\frac{d\beta}{dr}\right)^2
\eeq denotes the relativistic shear flow coefficient, with $\gamma_b(r)^2=1/(1-\beta(r)^2)$
\citep{Rieger2006,Webb2018}. Formally, Equation~(\ref{pte}) is a mixed-frame particle transport equation 
where the momentum variable $p$ is evaluated in the comoving flow frame, while the spatial  
coordinate $r$ is measured in the laboratory frame. This utilizes the fact that the scattering process 
is most conveniently evaluated in the local fluid frame. 
In general, the scattering time $\tau$ can be a function of space and momentum. In this paper  we 
focus on the $r$-independent case $\tau \equiv \tau(p) = \tau_0 p^{\alpha}$, $\alpha>0$. With this choice, 
Equation~(\ref{pte}) becomes a separable, elliptical partial differential equation. In quasi-linear theory $\alpha$ 
is related to the particle mean free path $\lambda = c \tau \propto p^{2-q}$ where $\alpha=2-q$ and $q$ 
is the power spectrum of the turbulence with $q = 1$ for Bohm, $q = 3/2$ for Kraichnan, and $q = 5/3$ for
Kolmogorov \citep[][]{Liu2017}.

\subsection{A Comparison with Analytical Results by Webb et al. (2018)}
\begin{figure*}[h]
\begin{center}
\includegraphics[width = 0.80 \textwidth]{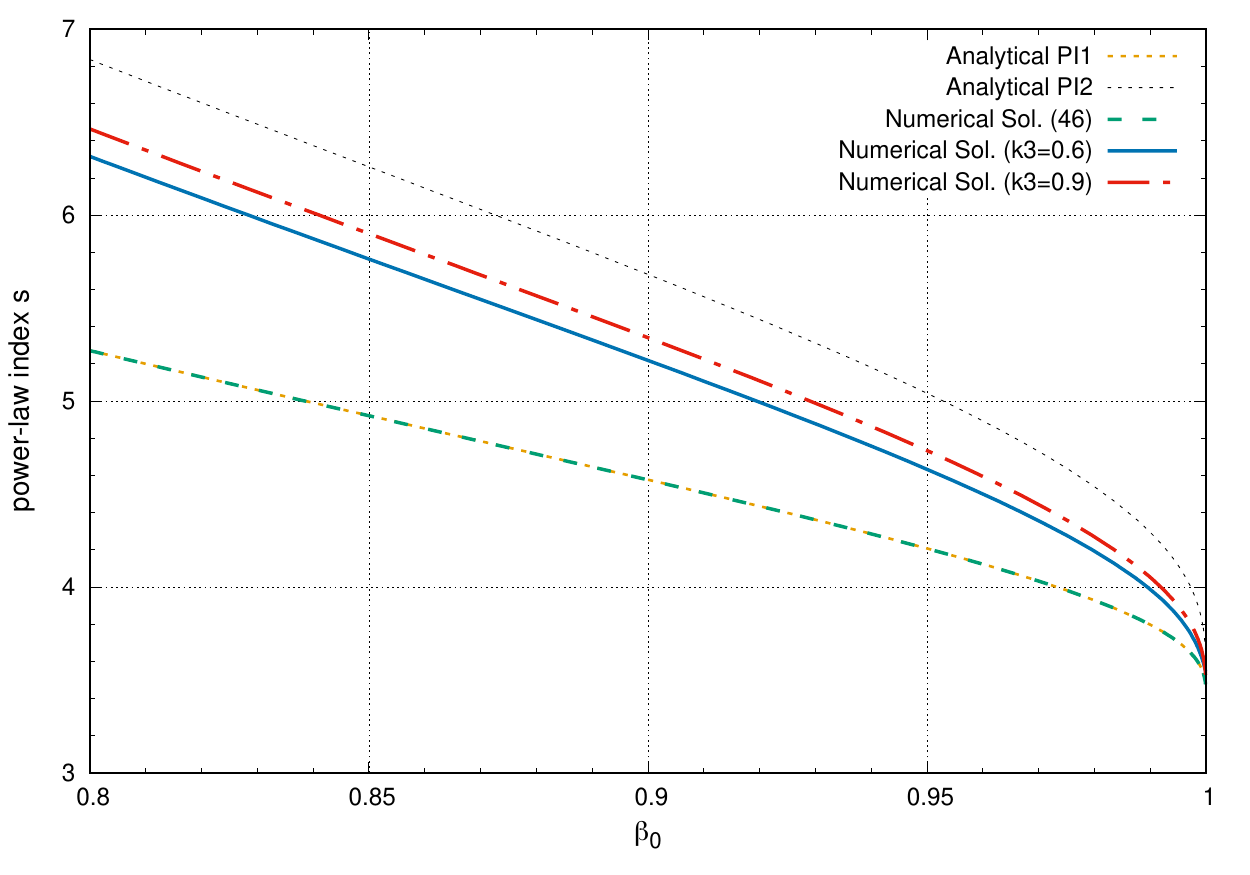}
\caption{Evolution of the momentum power-law index $s$, where $f\propto p^{-s}$, as a function 
  of the flow speed on the jet axis, $\beta_0\equiv \beta(r=0)$, for a particle mean free path 
  $\lambda =c\tau \propto p^{\alpha}$ with $\alpha=1/3$ (corresponding to Kolmogorov-type 
  turbulence $q=5/3$). The dotted orange line (PI1) denotes the asymptotic spectral index as 
  given in Equation~(\ref{pli1}) (see also \citet{Webb2018}), and the dashed green line (on top) 
  denotes the power-law index based on a numerical solution of the corresponding particle transport 
  equation \citep[][Equation~46]{Webb2018}. 
  Both agree, and they characterize the momentum spectrum in the case where $\tau$ shows a weak 
  $r$-dependence throughout most of the jet but has $\tau\rightarrow \infty$ as $r\rightarrow 0$. 
  The solid blue line denotes the numerical solution for a velocity profile with $k_3=0.6$ in
  Equation~(\ref{profile_Webb2018}), while the dashed-dotted red line denotes the one for $k_3=0.9$. 
  For both cases an $r$-independent $\tau$ has been assumed. The dotted black line at the top 
  gives the analytical asymptotic index as derived in \citet{Webb2019} for an $r$-independent $\tau$
  but a different nonsimilar flow profile (see main text below). For all cases, monoenergetic 
  injection has been considered, and $\beta_2=0$ has been assumed at the outer jet boundary.} 
\label{fig1}
\end{center}
\end{figure*}

\begin{figure*}[h]
\begin{center}
\includegraphics[width = 0.49 \textwidth]{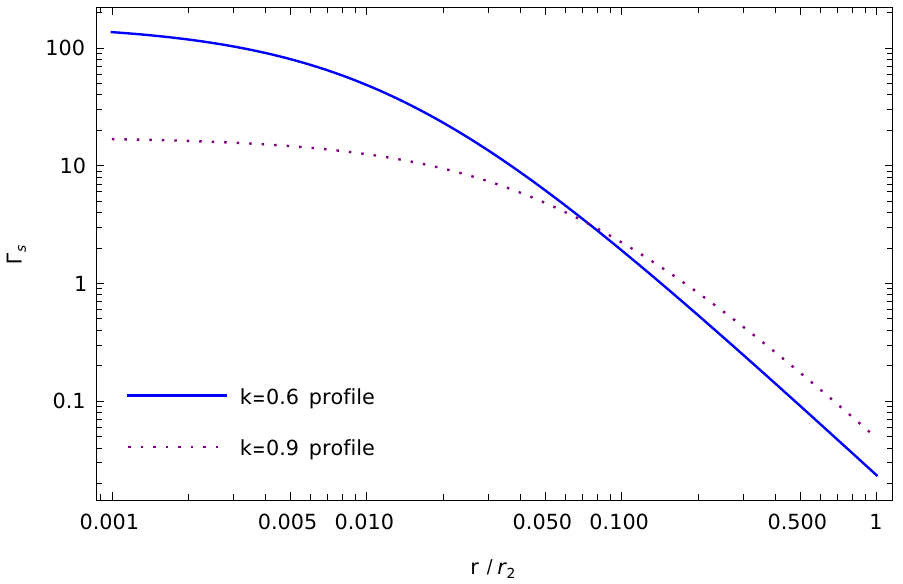}
\includegraphics[width = 0.49 \textwidth]{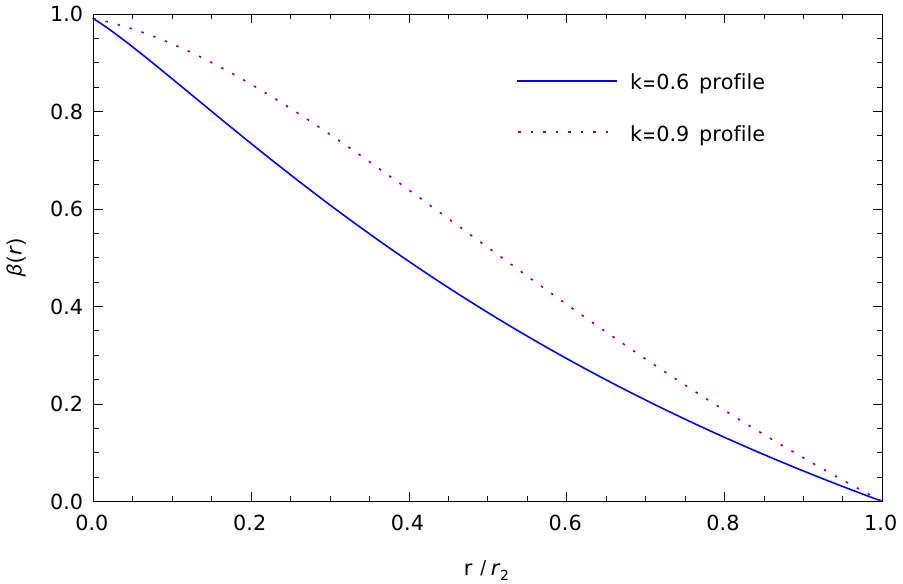}
\caption{{\it Left:} evolution of the shear coefficient $\Gamma_s$, Equation~(\ref{GammaS}), as a function 
  of the normalized radial coordinate $r/r_2$ for the velocity profiles explored in Figure~\ref{fig1}, see
  Equation~(\ref{profile_Webb2018}) and plot to the right, with $\beta_0=0.99$. The solid line corresponds to 
  the profile with $k_3 = 0.6$, while the dotted curve corresponds to the one with $k_3=0.9$. In both 
  cases, $\Gamma_s$ takes on large values but remains bounded as $r\rightarrow 0$, while 
  approximately showing a $\propto r^{-2}$-dependence on larger scales.
  {\it Right:} corresponding velocity profiles as function of the normalized radial coordinate $r/r_2$, 
  with line style identification as to the left (solid: $k_3 = 0.6$; dotted: $k_3=0.9$).} 
\label{fig2}
\end{center}
\end{figure*}

By solving the particle transport equation, the dependence of $f(r,p)$ on the flow speed $\beta$ can 
be explored. In particular, \citet{Webb2018} have presented results of Equation~(\ref{pte}) for the special 
case where $\tau(r,p) \propto p^{\alpha}/[\gamma(r)^2 r (d\beta/dr)]$, chosen as to yield analytical 
solutions. Requiring the scattering time $\tau$ to be nonsingular throughout most of the jet, this 
constraint can be related to some specific flow velocity profiles. For their model ansatz Equation~(36), for 
example, the resultant $\tau$ is weakly dependent on $r$, $\tau \propto  (1+\epsilon/r)$ where 
$\epsilon/r\ll1$, but has the property $\tau \rightarrow \infty$ as $r\rightarrow 0$. Since this property 
is restricted to a small region near the jet axis ($r=0$), this does not necessarily have to lead to unphysical 
results. However, since $\Gamma_s$ is very large in this region, a straightforward comparison with 
observations may not be possible. With boundary conditions $\partial f/\partial r \rightarrow 0$ as 
$r \rightarrow 0$ and $f(r_2,p) =0$ at the outer jet boundary $r_2$, \citet{Webb2018} showed that 
the particle energy distribution at high energies approaches a power-law particle distribution 
$f(r,p) \propto p^{-s}$ with index $s$ given by 
\beq\label{pli1}
s  =  \frac{3+\alpha}{2} + \sqrt{ \frac{(3+\alpha)^2}{4} 
    +  5 \pi^2 \left[ \ln \left(\frac{1+\beta_{02}}{1-\beta_{02}}\right) \right]^{-2} }\,,
\eeq where $\beta_{02} = (\beta_0-\beta_2)/(1-\beta_0\beta_2)$ is the relativistic relative velocity of 
the central jet velocity, $\beta_0$, to the outer velocity of the jet, $\beta_2=\beta(r_2)$. Obviously, as 
$\beta_0$ becomes nonrelativistic, $s$ increases significantly, implying very soft spectra. To explore 
consequences related to the singular $\tau$-description, we employ a velocity profile following 
\citet{Webb2018}, 
\beq\label{profile_Webb2018}
\beta(r) = \frac{(1+\beta_0)/(1-\beta_0)-[(1+r/\epsilon)]^{2 k_3}}{(1+\beta_0)/(1-\beta_0)
+[(1+r/\epsilon)]^{2 k_3}}\,, 
\eeq but assume $\tau$ to be $r$-independent, i.e. $\tau \equiv \tau(p) = \tau_0 p^{\alpha}$. 
Here and in the following we choose a Kolmogorov-type turbulence ($q=5/3$, i.e., $\alpha=1/3$). 
The flow profile of Equation~(\ref{profile_Webb2018}) takes on a maximum value $\beta_0$ at $r=0$;
$\epsilon \ll r_2$ is constrained by the requirement that $\beta(r_2) = 0$ (see Fig~\ref{fig2}).
Figure~\ref{fig1} shows examples of the evolution of the momentum spectral index $s$ above injection as 
a function of jet flow speed on the axis, $\beta_0$, based on a numerical solution of Equation~(\ref{pte}) with 
a finite element method using the default Pardiso direct solver \citep{Wolfram2022}. 
While all spectra approach the limiting value $s\rightarrow(3+\alpha)$ for ultrarelativistic speeds
$\beta_0\rightarrow 1$, there is a general trend for the momentum spectra to be softer than those 
based on a $\tau$-description that becomes singular as $r \rightarrow 0$. While the power-law 
index in Equation~(\ref{pli1}) is seemingly independent of the velocity shape, through the value of $k_3$, 
this no longer applies to the $r$-independent case. This seems partly due to the fact that the average 
rate of momentum change for a particle, $\langle\Delta p/\Delta t \rangle$, is proportional to 
$\Gamma_s p \tau$ \citep{Rieger2006,Liu2017}. Since, for relativistic $\beta_0$, $\Gamma_s$ takes 
on large values on small scales (see Figure~\ref{fig2}), this is expected to dominate particle energization, 
in particular for the considered singular $\tau$-description, thus leading to an apparent universal index 
evolution. We thus consider $s$, as defined in Equation~(\ref{pli1}), as giving a lower limit for velocity 
profiles of the type given by Equation~(\ref{profile_Webb2018}). 

\subsection{A Comparison of Different Flow Profiles for Spatially Constant $\tau$}
For a spatially constant $\tau$, on the other hand, \citet{Webb2019} have identified a class of 
concave-type velocity profiles, 
\beq\label{Webb2019}
\beta(r) = \tanh\left[\xi_0-(\xi_0-\xi_2)\left(\frac{r}{r_2}\right)^k \right],\; k\geq 1\,,
\eeq with $\xi_0=\mathrm{arctanh}(\beta_0)$ and $\xi_2=\mathrm{arctanh}(\beta_2)$, that give  
analytical solutions of the particle transport equation~(\ref{pte}). Independently of the value of $k$ 
(i.e., encompassing a variety of different shapes), they find an asymptotic power-law index for this 
case given by 
\beq\label{pli2}
s  =  \frac{3+\alpha}{2} + \sqrt{ \frac{(3+\alpha)^2}{4} 
    +  20 j_{0,1}^2 \left[ \ln \left(\frac{1+\beta_{02}}{1-\beta_{02}}\right) \right]^{-2} }\,,
\eeq where $j_{0,1}\simeq 2.4048$ denotes the first zero of the $J_0(z)$ Bessel function of the first 
kind \citep[][]{Abram1972}.
We find very good agreement between a numerical solution of Equation~(\ref{pte}) for the particular case 
of $k=2$ and this analytically derived expression. The power-law index, Equation~(\ref{pli2}), is comparable 
to, but softer than, the one in the previously considered case (where $\tau\rightarrow\infty$ as 
$r\rightarrow 0$), Equation~(\ref{pli1}), and may suggest that for spatially constant diffusion softer 
spectra are obtained. To investigate this in more detail, we numerically solve the particle transport 
equation, Equation~(\ref{pte}), for three different velocity class profiles: the 
linear case 
\beq\label{linear}
 \beta_l(r) = \beta_0 \left(1- (r/r_2)\right)\,,
\eeq a Gaussian profile, with $a_g>0$, 
\beq\label{Gaussian}
 \beta_g(r) = \beta_0 \exp\left(-a_g\,[r/r_2]^2\right)\,,
\eeq 
and a  power-law dependence, with $b\ge 1$, 
\beq\label{power-law}
\beta_p(r) = \frac{\beta_0}{(1+a_p^2\, [r/r_2]^2)^{b/2}}\,.
\eeq 
The Gaussian and power-law profiles are convex over most of the interval and in that respect different 
from the concave profile of Equation~(\ref{Webb2019}). For flow profiles of the type given by 
Equation~(\ref{profile_Webb2018}), $\Gamma_s$ is a monotonically decreasing function, with maximum 
values on the axis, while for flow profiles of the type given by Equation~(\ref{Webb2019}), $\Gamma_s$ 
is a monotonically increasing function (for $k>1$) that takes on a maximum value at the outer boundary. 
On the other hand, the linearly decreasing profile, Equation~(\ref{linear}), results in a monotonically 
decreasing $\Gamma_s$, while profiles of the type given by Equations~(\ref{Gaussian}-\ref{power-law}) 
result in a $\Gamma_s$, that initially rises and peaks close to, but not on, the axis and then subsequently 
decreases monotonically (see Figure~7 in the Appendix).
\begin{figure}[h]
\begin{center}
\includegraphics[width = 0.80 \textwidth]{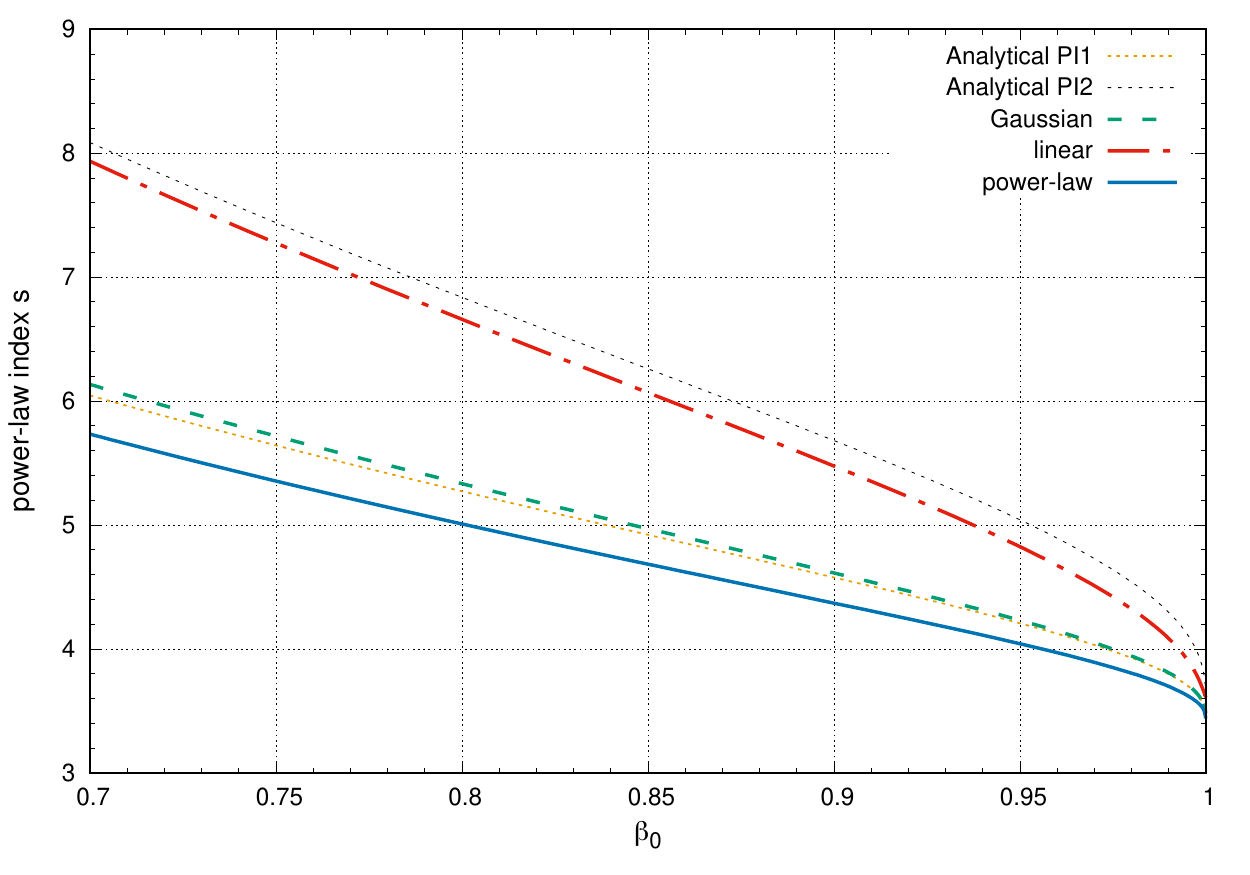}
\caption{Evolution of the momentum power-law index $s$, where $f \propto p^{-s}$, as a function of 
  the flow speed on the jet axis, $\beta_0 \equiv \beta(r=0)$, for spatially constant $\tau \propto 
  p^{\alpha}$ with $\alpha=1/3$. 
  The dotted blue line (PI1) denotes the asymptotic spectral index as given in Equation~(\ref{pli1}) 
  \citep{Webb2018}, and the upper, dotted black line (PI2) the asymptotic index as given in 
  Equation~(\ref{pli2}) \citep{Webb2019}, both considering $\beta_2=0$. The dashed-dotted red line 
  denotes the numerical solution for a linearly decreasing velocity profile, Equation~(\ref{linear}), 
  and the dashed green line the one for a Gaussian velocity profile, Equation~(\ref{Gaussian}), with 
  $a_g=5$. While the index for a linear decreasing velocity profile is relatively soft and close 
  to the one for a concave profile, Equation~(\ref{Webb2019}), the one for a Gaussian velocity profile 
  is relatively hard and close to the singular $\tau$ case. Even harder momentum spectra are 
  obtained, however, for a power-law-type velocity profile, Equation~(\ref{power-law}), with $a_p=5$, 
  $b=2$ (solid, purple line).}
\label{fig3}
\end{center}
\end{figure}
The numerical results for a monoenergetic source term with $p=p_0$ at $r=0.1\,r_2$ are shown in 
Figure~\ref{fig3}. As before, boundary conditions $\partial f/\partial r \rightarrow 0$ as $r \rightarrow 0$ 
and $f(r_2,p) =0$ at the outer jet boundary $r_2$ have been applied. At high energies the particle
distribution $f(r,p)$, evaluated at $r=0.8\,r_2$, follows a power law with somewhat different index
evolution. For the linearly decreasing velocity profile the spectra are relatively soft, with power-law 
index $s$ close to the one for the concave profile of Equation~(\ref{Webb2019}) and to the one for a 
$k_3=0.9$ profile, Equation~(\ref{profile_Webb2018}) (see Figure~\ref{fig1}), while the spectra for 
a Gaussian velocity profile are relatively hard, with index very close to the one for the singular $\tau$ 
case, Equation~(\ref{pli1}). 
The hardest spectrum (e.g., at $\beta_0 = 0.95$ by $\Delta s \sim 0.2$ when compared to the 
singular one) is obtained for a power-law-type velocity profile, Equation~(\ref{power-law}). The latter 
result provides evidence that even for spatially constant diffusion, momentum spectra harder than 
the asymptotic index, Equation~(\ref{pli1}), can be obtained. At ultrarelativistic speeds, we observe 
$s \rightarrow (3+\alpha)$ for all profiles.

\subsection{Solutions for a Power-law-type Velocity Profile}
\begin{figure}[h]
\begin{center}
\includegraphics[width = 0.46 \textwidth]{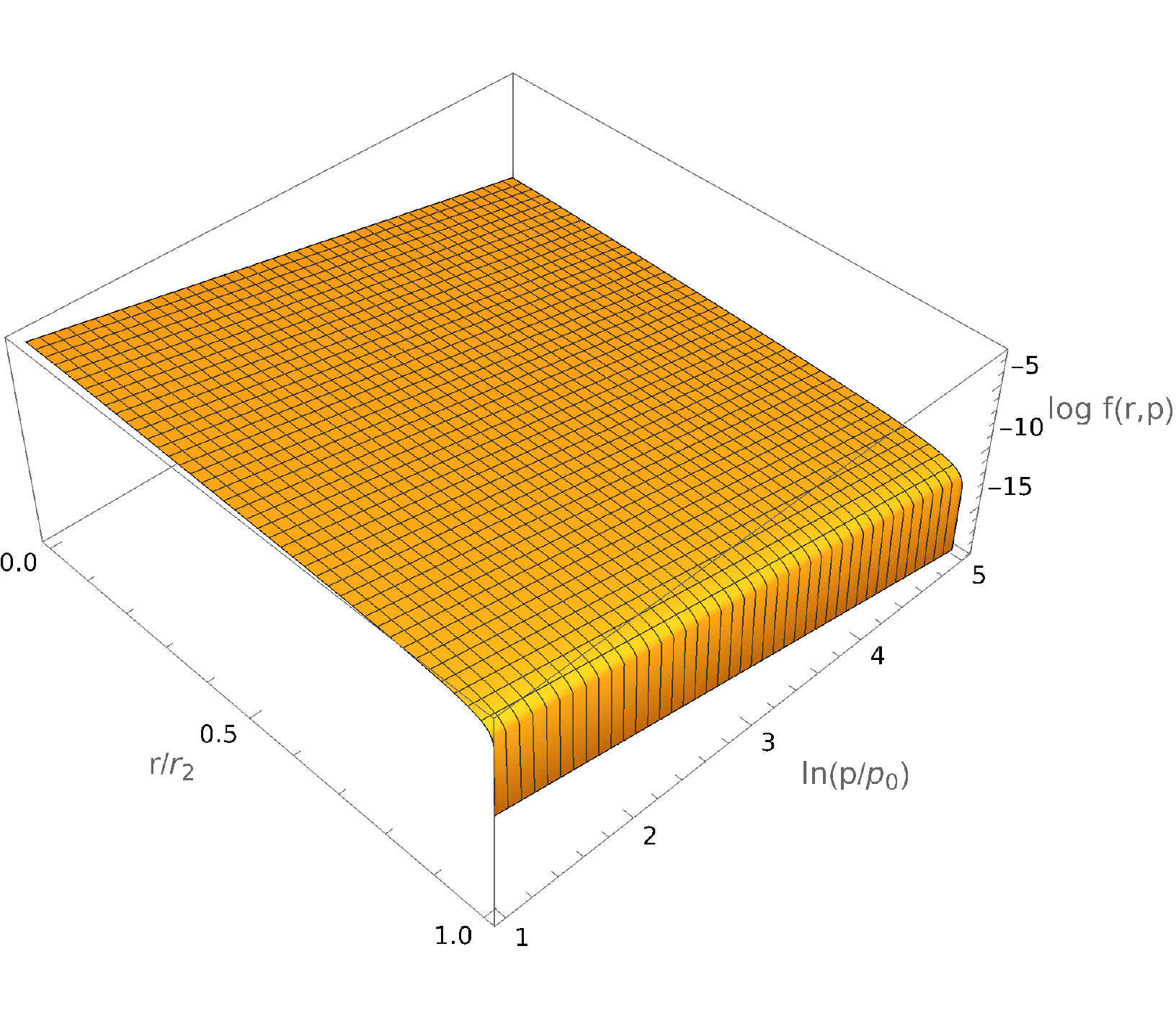}
\includegraphics[width = 0.51 \textwidth]{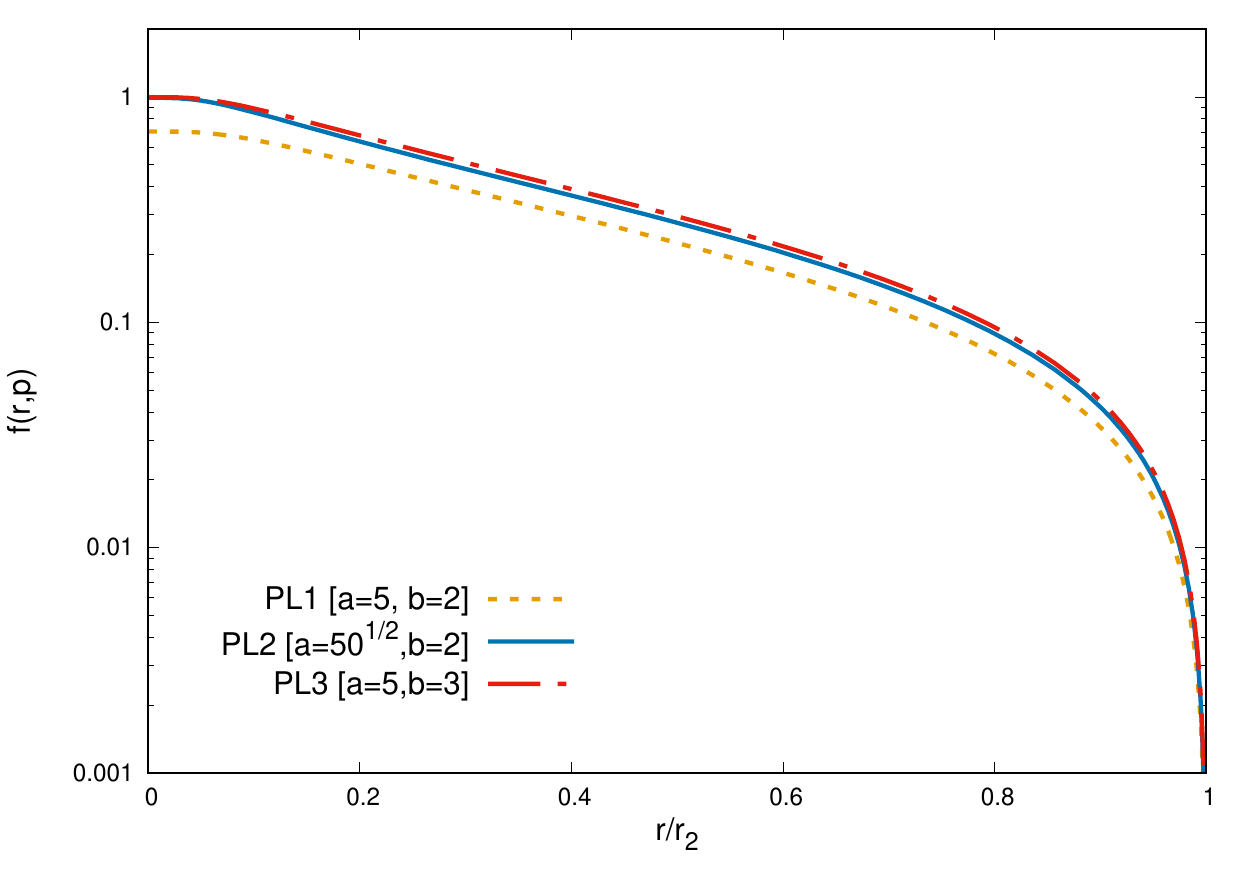}
\caption{{\it Left:} three-dimensional plot of $f(r,p)$ for a power-law-type velocity profile, 
  Equation~(\ref{power-law}), with $a_p=5$, $b=2$ and a flow speed on the axis of $\beta_0=0.95$. 
  {\it Right:} exemplary illustration of the $r$-dependence of $f(r,p)$ at fixed $p=e^3\,p_0$ 
  for power-law velocity profiles, Equation~(\ref{power-law}), with $a_p=5, b=2$ (dotted orange line), 
  $a_p=\sqrt{50}, b=2$ (solid blue line), and $a=5, b=3$ (dashed-dotted red line), all particle 
  distribution functions being normalized to the value of $f$ at $r=0$ for the third profile. A 
  flow speed on the axis of $\beta_0=0.90$ has been employed.} 
\label{fig4}
\end{center}
\end{figure}

\begin{figure}[htbp]
\begin{center}
\includegraphics[width = 0.75 \textwidth]{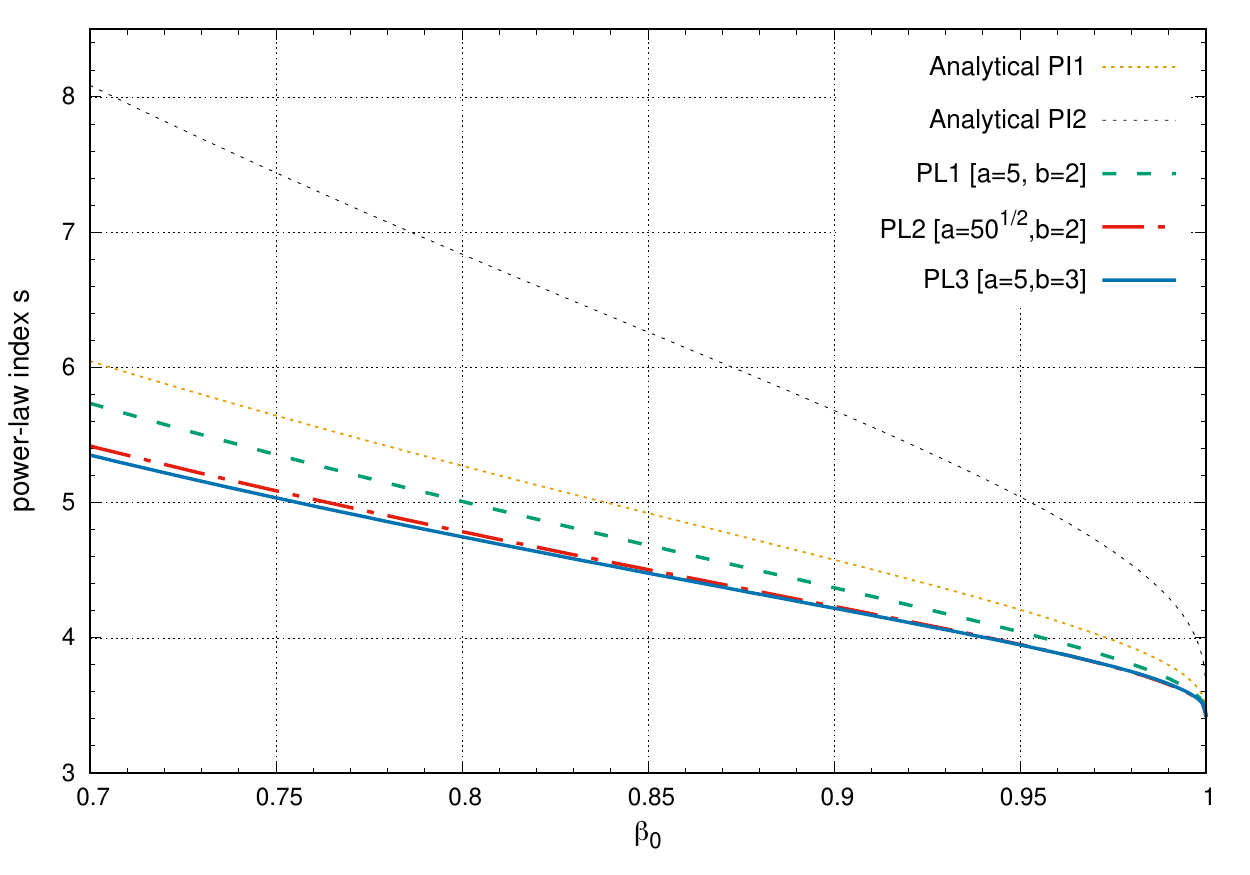}
\caption{Evolution of the momentum power-law index $s$, where $f \propto p^{-s}$, as a function 
  of the flow speed on the jet axis, $\beta_0 \equiv \beta(r=0)$, for spatially constant $\tau 
  \propto p^{\alpha}$ with $\alpha =1/3$. The dashed orange line (PI1) denotes the asymptotic 
  spectral index as given in Equation~(\ref{pli1}) \citep{Webb2018}, and the dotted black line (PI2) 
  denotes the asymptotic index as given in Equation~(\ref{pli2}) \citep{Webb2019}, both considering 
  $\beta_2=0$. The three other lines denote the numerical solution for a power-law profile, 
  Equation~(\ref{power-law}), with  $a_p=5, b=2$ (dashed green), $a_p=\sqrt{50}, b=2$ (dashed-dotted 
  red) and $a_p=5, b=3$ (solid blue line).}
\label{fig5}
\end{center}
\end{figure}

As the power-law velocity profile yields the hardest spectra, we further explore two more cases, 
varying $a_p$ and/or $b$ in Equation~(\ref{power-law}). We consider, as before, monoenergetic 
injection with $p=p_0$ at $r=0.1\,r_0$, and $\alpha=1/3$. 
Figure~\ref{fig4} (left) shows a three-dimensional plot of $\log f(r,p)$ for the previous case 
$a_p=5, b=2$ over the spatial range $r=(0-1)\,r_s$ and a momentum range $p$ above injection. 
Figure~\ref{fig4} (right) shows the $r$-dependence of the respective $f(r,p)$ for the three different 
power-law-type profiles at fixed momentum $p$, with the distribution functions appearing 
comparatively similar and satisfying the imposed boundary conditions. 
The dependence of $s$ on $\beta_0$ is shown in Figure~\ref{fig5} which becomes harder for steeper 
profiles, i.e. for $a_p=\sqrt{50}, b=2$ (red line) and $a_p=5, b=3$ (purple line). The indices for the 
latter cases appear close to each other, as their velocity shapes and shear coefficients appear 
relatively similar. In particular, power-law momentum indices $s\simeq 4$, corresponding to $N(E) 
\propto E^{-2}$, are achievable for flow Lorentz factors $\Gamma_0 \simeq 3$ ($\beta_0 \simeq 0.94$).

\section{Comparison with the Leaky-box Spectrum}
The above results illustrate that in general the power-law momentum index $s$ depends on the 
shape of the bulk flow profile. In principle, this is to be expected, as particle energization satisfies 
$\langle \Delta p/\Delta t \rangle \propto \Gamma_s p \tau$.
In this section we explore the extent to which these results can be recaptured by means of the 
simplified leaky-box Equation (\ref{diffusion}), where $D_p$ is the momentum-space shear 
diffusion coefficient \citep{Rieger2006} $D_p = \Gamma_s p^2 \tau =: D_0\, p^{2+\alpha}\,$, with 
$\tau=\lambda/c \propto p^{\alpha}$ the (momentum-dependent) mean scattering time and 
$\Gamma_s$ the shear coefficient as defined in Equation~(\ref{GammaS}). The leaky-box approach 
assumes that all spatial dependencies, for example, the second term in Equation~(\ref{pte}), can be 
approximated by some suitably chosen average escape term. 
Following Equation~(\ref{diffusion}), the characteristic (comoving) shear particle acceleration timescale 
might be expressed as \citep[e.g.,][]{Rieger2019}
\beq\label{tacc}
 t_{\rm acc}(p) = \frac{c}{(4+\alpha)\, \Gamma_s \,\lambda} \propto p^{-\alpha}\,.
\eeq The escape time, on the other hand, is determined by cross-field transport and the width of the jet
\beq\label{tau_esc} 
 \tau_{\rm esc}(p) = \frac{(\Delta r)^2}{2\,\kappa(p)}\propto p^{-\alpha} \,,
\eeq where $\kappa=\lambda c/3$ denotes the spatial diffusion coefficient and $\Delta r=(r_2-r_0)$ 
\citep[][]{Rieger2019b} where we take $r_0=0$.\\
The Green's function solution of Equation~(\ref{diffusion}) for monoenergetic injection $Q_{\delta}(p) 
= Q_0\,\delta(p-p_0)$ is given by \citep[see][]{Rieger2019b,Webb2020}
 \beq
 f(p) = f_0\, p^{-s_1}\, H(p_0-p) + f_0 \, p^{-s_2}\, H(p-p_0)\,,
\eeq where the corresponding power-law indices are given by
\beq\label{PI_LB}
 s_{1,2} = \frac{(3+\alpha)}{2} \mp \sqrt{\frac{(3+\alpha)^2}{4} + (4+\alpha)\, 
            \frac{t_{\rm acc}}{\tau_{\rm esc}}}
        =  \frac{(3+\alpha)}{2} \mp \sqrt{\frac{(3+\alpha)^2}{4} + \frac{2\,c^2}{3\, \Gamma_s r_2^2}}\,
 \,.
\eeq For $t_{\rm acc} \ll \tau_{\rm esc}$, i.e. for relativistic shear flows, $s_1 \rightarrow 0$ 
and $s_2 \rightarrow (3+\alpha) = 10/3$ for $\alpha=1/3$. For subrelativistic flows, on the other 
hand, the second term inside the square root becomes large, resulting in steep (soft) high-energy 
spectra.\\
However, even for an $r$-independent scattering time $\tau$, the power-law expression in 
Equation~(\ref{PI_LB}) formally includes an explicit $r$-dependence (via $\Gamma_s$ in $t_{\rm acc}$, 
see Equations~[\ref{tacc}] and [\ref{GammaS}]). In \citet{Rieger2019b} we explored a possible way of 
treating this $r$-dependence, approximating $\gamma_b(r)^4 (d \beta/d r)^2$ in $\Gamma_s$, 
Equation~(\ref{GammaS}), by $\langle \gamma_b(r)^2\, (d\beta/dr) \rangle^2$, where $\langle \rangle$ 
denotes averaging over $r$ from $0$ to $r_2$. Since for $\beta(0)=\beta_0$ and $\beta(r_2) =
\beta_2$, $\langle \gamma_b(r)^2 (d\beta/dr) \rangle = -(1/[2\,r_2]) 
\ln([1+\beta_{02}]/[1-\beta_{02}])$, where $\beta_{02} = (\beta_0-\beta_2)/(1-\beta_0\beta_2)$, 
this approximation then yields
\beq
 s  =  \frac{(3+\alpha)}{2} 
     + \sqrt{\frac{(3+\alpha)^2}{4} + 40\,\left[\ln\frac{(1+\beta_{02})}{(1-\beta_{02})}\right]^{-2}}\,,
\eeq for the particle distribution above injection (i.e., for $p> p_0$) in the leaky-box approach.
This expression turns out to be close to the analytical result of Equation~(\ref{pli1}), reproducing 
well the qualitative behavior of $s$ as a function of the on-axis flow speed $\beta_0$ 
\citep[see also][]{Webb2020}. 
One obvious drawback of this approximation is, however, that it does not allow one to distinguish 
between different velocity shapes since the resultant expression is only a function of the on-axis 
$\beta_0$ and the outer $\beta_2$. 

To improve on this in view of our current results, we use the original particle transport 
equation~(\ref{pte}) with a local, monoenergetic source term, applying separation of variables, 
$f(r,p) = h(r)\cdot g(p)$. Away from injection and assuming an $r$-independent $\tau$, one 
then obtains the two differential equations,
\beq\label{sep}
\frac{1}{\Gamma_s} \frac{1}{r} \frac{\partial}{\partial r} 
\left(r \frac{\partial h(r)}{\partial r}\right) = - \lambda^2\cdot h(r)\,,\;\;\;
\frac{1}{\kappa\,p^2}\frac{\partial}{\partial p} 
\left(p^4 \tau \frac{\partial g(p)}{\partial p}\right) = \lambda^2 \cdot g(p)\,,
\eeq where $\lambda^2$ is a constant to be determined by the boundary conditions. Replacing 
$\Gamma_s$ by a suitable, spatially averaged $\langle\Gamma_s\rangle=:\tilde{\Gamma}_s/r_2^2$, 
then the first differential equation can be solved and has the general solution 
\beq\label{sol}
  h(r) = c_1 J_0\left (\lambda \sqrt{\tilde{\Gamma}_s}\,\frac{r}{r_2}\right) 
         + c_2 Y_0\left (\lambda \sqrt{\tilde{\Gamma}_s}\, \frac{r}{r_2}\right)\,,
\eeq where $c_1, c_2$ are constants, and where $J_0$ and $Y_0$ denote the Bessel functions 
of the first and second kind, respectively \citep[][]{Abram1972}. The boundary conditions 
$\partial h/\partial r \rightarrow 0$ as $r \rightarrow 0$ and $h(r_2)=0$, then require $c_2=0$, 
noting that $J_0'(x)=-J_1(x)$, and $Y_0'(x) = -Y_1(x)$, and $\lambda_n = j_{0,n} /\sqrt{\tilde{\Gamma}_s}$ 
where $\lambda_n$ is the $n$th zero ($n=1,2,...$) of the Bessel function $J_0$. For $n=1$, a 
fundamental solution is thus fixed, implying $\lambda=\lambda_1= -j_{0,1}/\sqrt{\tilde{\Gamma}_s}$, 
where $j_{0,1} = 2.4048$. With $g(p) \propto p^{-s}$ in the second differential equation~(\ref{sep}) one 
obtains $s_{1,2} = (3+\alpha)/2 \pm \sqrt{(3+\alpha)^2/4 + c^2 \lambda^2/3}$, so that with the noted 
$\lambda$ one obtains
\beq\label{s_approx}
s  =  \frac{3+\alpha}{2} + \sqrt{ \frac{(3+\alpha)^2}{4} +  \frac{c^2}{3} 
\frac{j_{0,1}^2}{\tilde{\Gamma}_s}}
\eeq for the power-law momentum index of the particle distribution above injection. 

\begin{figure}[htbp]
\begin{center}
\includegraphics[width = 0.80 \textwidth]{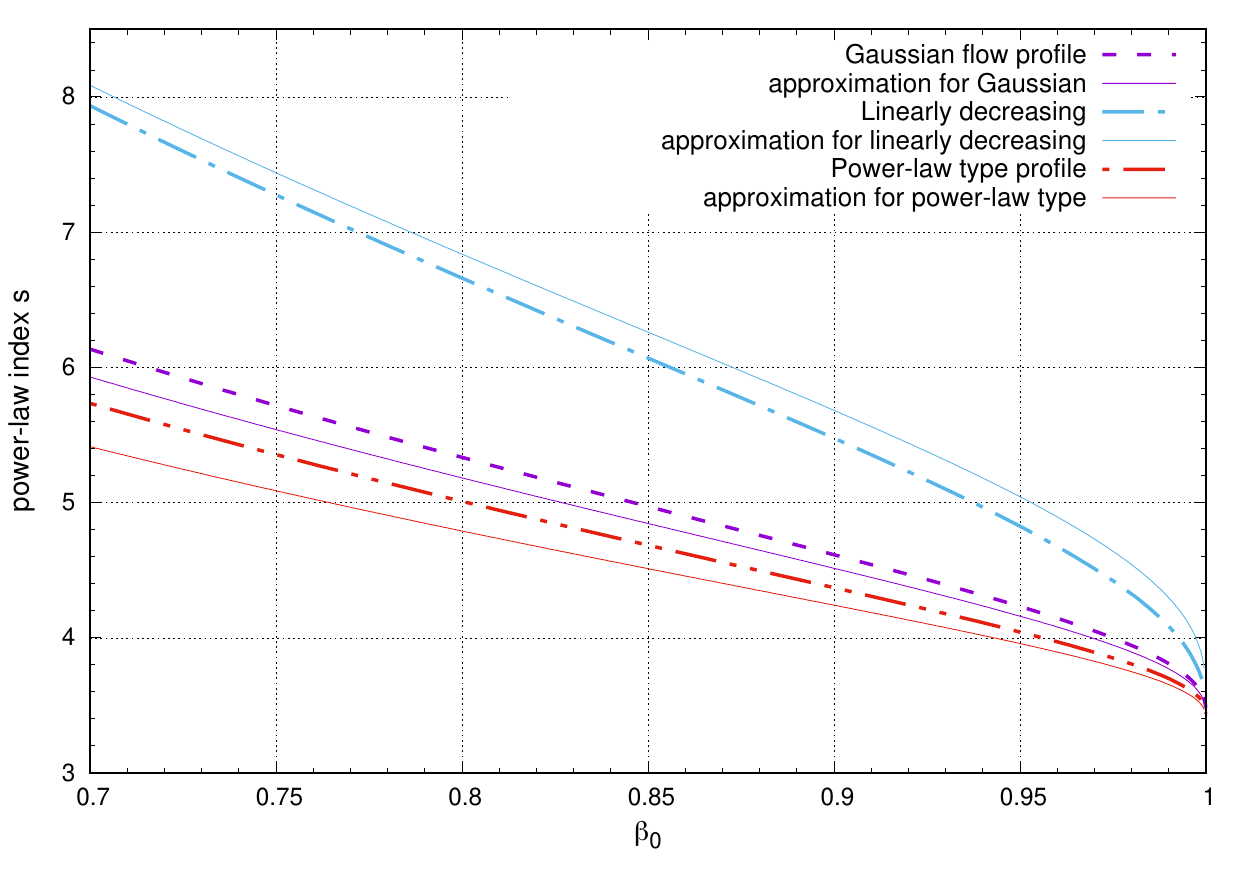}
\caption{Evolution of the momentum power-law index $s$ as a function of the flow speed on 
  the jet axis, $\beta_0 \equiv \beta(r=0)$, for spatially constant $\tau \propto p^{\alpha}$ with
  $\alpha =1/3$. The thick solid lines denote numerical solutions of the particle transport 
  equation~(\ref{pte}) for a linearly decreasing, Gaussian and power-law-type ($a_p=5, b=2$) 
  profile, respectively, as in Figure~\ref{fig3}. Thin solid lines represent approximations (see 
  text and Equation~[\ref{approx}] for details) based on a leaky-box approach. These 
  approximations deviate from the exact results by less than $\sim6\%$.}
\label{fig6}
\end{center}
\end{figure}

For the special case of a linearly decreasing profile, where $\Gamma_s$ is a monotonically 
decreasing function, our previous choice of averaging \citep{Rieger2019b} turns out to work
rather well. When the corresponding $\tilde{\Gamma}_s= (c^2/60)\,\left(\ln\left[(1+\beta_{02})/
(1-\beta_{02})\right]\right)^2$ is inserted in Equation~(\ref{s_approx}) a reasonable approximation 
of the numerical solution is achieved (see Figure~\ref{fig6}). The approximation in this case 
actually coincides with Equation~(\ref{pli2}) \citep[see][]{Webb2019}. 
On the other hand, for a Gaussian and a power-law-type profile, for which $\Gamma_s$ peaks 
on small scales (close to the axis), we find that a quasi-weighted average 
\beq\label{approx}
 \tilde{\Gamma}_s = \frac{c^2}{15}\,
 \frac{r_2^2 \langle \gamma_b(r)^2 (d\beta/dr) \rangle^2}{\langle \beta(r) \rangle}
 = \frac{c^2}{60\,\langle \beta(r)\rangle}\, 
\left( \ln\left[\frac{1+\beta_{02}}{1-\beta_{02}}\right]\right)^2
\eeq with $\langle \beta(r) \rangle \equiv (\int_0^{r_2} \beta(r) dr) /(\int_0^{r_2} dr)$ yields 
a reasonable approximation when inserted in Equation~(\ref{s_approx}); see Figure~\ref{fig6}. 
While approximate, using expression Equation~(\ref{approx}) in Equation~(\ref{s_approx}) thus 
provides a simple means of taking the shape of the velocity profile into account and improving 
the estimate for the momentum spectral index based on a leaky-box approach. This corresponds 
to replacing $\tau_{\rm esc}$ in the previous leaky-box equation *(see Equations~(\ref{diffusion}), 
(\ref{tau_esc}), (\ref{PI_LB})) by $\tau_{\rm esc} \rightarrow \tilde{\tau} \,\tau_{\rm esc}$, 
where $\tilde{\tau}=2/j_{01}^2$.
The results above can in principle be used to facilitate a simple first-order comparison with 
observations.

\section{Discussion and Conclusion}
The results presented here confirm the intuitive notion that, in general, both the magnitude of the 
flow velocity and its explicit shear profile have an impact on the momentum spectrum generated in 
gradual shear particle acceleration. Considering spatially constant diffusion, we have explored 
several characteristic shear flow profiles, from a linearly decreasing to an exponentially decaying 
(Gaussian-type) velocity shear. When compared with each other, power-law-type velocity profiles 
provide the hardest (flattest) spectra. 

In the absence of radiative losses, nonthermal shear particle acceleration results in power-law 
momentum spectra. At ultrarelativistic speeds ($\beta \rightarrow 1$), the momentum spectral 
index for the particle distribution $N(p) \propto p^2 f(p) \propto p^{2-s} \propto p^{-\tilde{q}}$ 
becomes independent of the shape of the velocity profile, $s \rightarrow (3+\alpha)$, 
corresponding to $\tilde{q} \rightarrow (1+\alpha)$, with $\alpha$ parameterizing the 
momentum dependence of the mean scattering time, $\tau = \lambda/c \propto p^{\alpha}$. On 
the other hand, for mildly relativistic speeds power-law momentum indices $s\simeq 4$,
corresponding to $N(p) \propto p^{-2}$, are achievable, for example, for power-law-type flow 
profiles with Lorentz factors $\Gamma_0 \simeq 3$ ($\beta_0 \simeq 0.94$) and $\alpha=1/3$.

We find that a simple leaky-box-type approach for relativistic shear acceleration captures the 
qualitative and quantitative (to within $\leq 10\%$) behavior of the momentum spectral index 
as a function of the on-axis speed $\beta_0$ for both Gaussian and power-law-type velocity 
profiles with 
\beq
s \simeq \frac{(3+\alpha)}{2} + \left[\frac{(3+\alpha)^2}{4} 
+ 116 \,\langle \beta \rangle\, \left( \ln\frac{1+\beta_{02}}{1-\beta_{02}}\right)^{-2}\right]^{1/2}
\eeq 
where 
$\langle \beta \rangle<1$ denotes a weighted, spatial average of the considered velocity 
profile, and $\beta_{02} = (\beta_0-\beta_2)/(1+\beta_0\beta_2)$ is the relative velocity 
between the on-axis flow speed $\beta_0$ and flow speed $\beta_2$ at the outer jet (shear) 
boundary.

The treatment presented here does not include radiative losses (e.g., synchrotron). 
The latter could lead to a pileup in the particle distribution at the momentum scale where 
acceleration balances losses \cite[e.g.,][]{Tavecchio2021,Wang2021}. Taking this constraint 
into account, our present findings are providing a useful guide to cosmic-ray or electron shear 
acceleration over the energy regime for which radiative losses might be neglected.
The current study is limited to spatially constant diffusion in order to facilitate a comparison 
of different flow profiles. In principle, an extended analysis keeping $\tau(r,p)$ is desirable, 
but it is left to a future paper. However, the results presented here are of relevance for 
the interpretation of jetted, high-energy astrophysical sources such as AGNs, since nonthermal 
shear acceleration is a promising mechanism for the acceleration of charged particles to 
relativistic energies and thus possibly responsible for some of the high-energy radiation 
observed \citep[][for a review]{Rieger2019}.  For example, the recently detected, extended 
very high energy emission along the kiloparsec-scale jet of Centaurus A has been successfully 
modeled as inverse Compton emission by a relativistic electron distribution whose high-energy 
branch follows a power law with index $\tilde{q}=3.85$ \citep{HESS2020}. This would be compatible 
with a Gaussian-type shear flow profile (provided $\beta_0\sim 0.7$) or a power-law-type one 
(for intermediate $\beta_0 \sim 0.5-0.6$). If one takes bulk velocity estimates based on radio 
observations as reliable indicators of the maximum jet speed, which may not necessarily be 
the case, then this would suggest $\beta\sim 0.5$ \citep[][]{Hardcastle2003} and tend to favor 
power-law-type velocity profiles over Gaussian ones. 

In conclusion, the results presented in this paper provide a foundation for relating the 
momentum power-law index inferred from spectral modeling to specific flow velocity 
profiles and establish a relationship between $\beta_0$ and a required power-law 
index $s$ (or $\tilde{q}$), both of which might be observationally constrained for 
specific sources.

\acknowledgments
F.M.R. kindly acknowledges funding by the DFG under RI 1187/8-1. 
 
\bibliography{references}{}
\bibliographystyle{aasjournal}

\appendix
\begin{figure}[htbp]
\begin{center}
\includegraphics[width = 0.80 \textwidth]{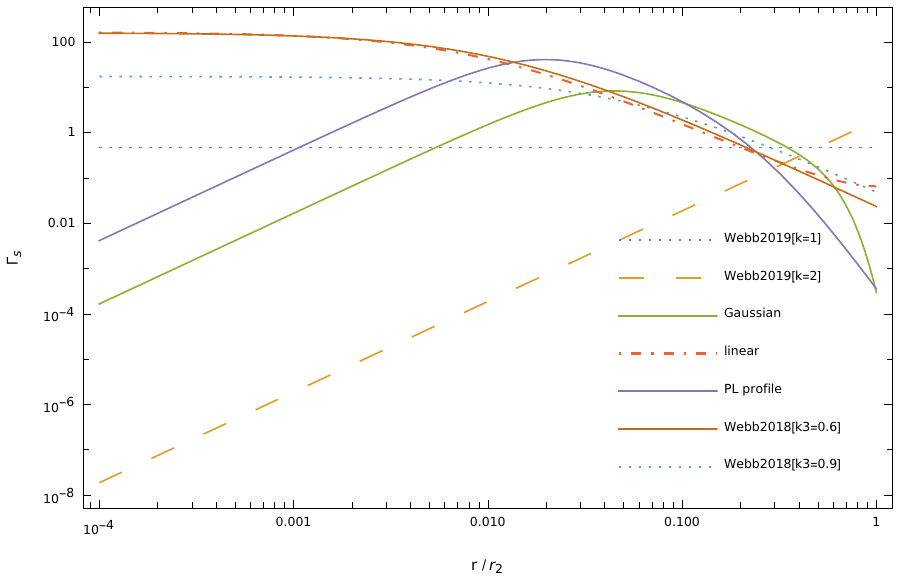}
\caption{Illustration of the relativistic shear flow coefficient $\Gamma_s$, Equation~(\ref{GammaS}), as a function 
 of the normalized radial coordinate $r/r_2$ for flow profiles considered in this work and evaluated assuming 
 $\beta_0 = 0.99$, i.e., Gaussian, Equation~(\ref{Gaussian}) with $a_g=5$; linearly decreasing, Equation~(\ref{linear}); 
 power-law-type (PL), Equation~(\ref{power-law}) with $a_p=5, b=2$; concave-type profiles following \citet{Webb2019}, 
 Equation~(\ref{Webb2019}), for $k=1$ and $k=2$; and flow profiles following \citet{Webb2018}, 
 Equation~(\ref{profile_Webb2018}), with $k_3=0.6$ and $k_3=0.9$, respectively. Gaussian and power-law-type
 profiles show a maximum (peak) on small spatial scales.}
\label{fig7}
\end{center}
\end{figure}



\end{document}